\documentclass[12pt]{article}
\usepackage{amsfonts}
\usepackage{mathrsfs}
\usepackage{latexsym,amsmath}
\usepackage{amssymb,array}
\usepackage{hyperref}
\makeatletter \oddsidemargin 0in \evensidemargin 0in \textwidth
16cm 
\RequirePackage[dvips]{graphicx}
\begin{document}
\newcommand{\wtl}{\widetilde}
\newcommand{\ov}{\overline}
\newcommand{\om}{\omega}
\newcommand{\ga}{\gamma}
\newcommand{\cd}{\circledast}
\newtheorem{thm}{Theorem}[section]
\newtheorem{remark}{Remark}[section]
\newtheorem{counterexample}{Counterexample}[section]
\newtheorem{coro}{Corollary}[section]
\newtheorem{definition}{Definition}[section]
\newtheorem{example}{Example}[section]
\newtheorem{lem}{Lemma}[section]
\numberwithin{equation}{section}
\title{General Standby Component Allocation \\in Series and Parallel Systems}
 \author{Nil Kamal Hazra and Asok K. Nanda \footnote{e-mail: asok.k.nanda@gmail.com, corresponding author.}
 \\Department of Mathematics and Statistics
  \\IISER Kolkata, Mohanpur Campus
 \\Mohanpur 741252, India
}
 \date{January, 2014}
\maketitle
\begin{abstract}
\hspace*{0.2 in}Optimum lifetime of a series (resp. parallel) system with general standby component(s) always depends on allocation strategy of standby component(s) into the system.
Here, we discuss three different models of one or more standby components.
In each model, we compare different series (resp. parallel) systems (which are formed through different allocation strategies of standby component(s)) with respect to 
the usual stochastic and the stochastic precedence orders.
\end{abstract}
$\mathbf{Key\,Words\,and\,Phrases}:$ Series system, standby system, stochastic orders, parallel system. 
\section{Introduction}
\hspace*{0.3 in}It is an eternal truth that every system must be collapsed after certain time. For this reason, reliability engineers show their keen interest to find out different ways
by which reliability of the system could be increased. Allocation of standby (or redundant) component(s) into the system is an effective way to enhance the lifetime of the system. Then, the 
natural question is $-$ how and where to allocate standby component(s) into the system so that system reliability will become optimum? In this note we have discussed three different
models which suggest some possible answers of this question.
Standby components are mostly of three types $-$ hot (or active) standby, cold standby and warm standby. In hot standby, the original component and the redundant component work together  
under the same operational environment. 
In cold standby, the redundant component has zero failure rate when it is in inactive state. 
It starts to function under the usual environment (in which the system is running) only when the original component fails.
On the other hand, warm standby describes an intermediate scenario. 
In warm standby, the redundant component undergoes two operational environments. 
Initially, it functions in a milder environment (in which a redundant component has less failure rate than its actual failure rate), there after it switches over 
to a usual environment after the original component fails.
It might happen that the redundant component fails before switching over to the usual environment.
Warm standby is sometimes called {\it general standby} because it contains both the hot standby and the cold standby as extreme cases. From now onwards, by warm standby we mean general standby. 
Both hot as well as cold standby allocation problems have been widely studied in the literature, for instance,
see, Boland {\it et al.}~\cite{bep7}, She and Pecht~\cite{sp7}, Singh and Misra~\cite{sm7}, Romera {\it et al.}~\cite{rvz2},
Li and Hu~\cite{lh22}, Brito {\it et al.}~\cite{br1}, Li {\it et al.}~\cite{lyh2}, Misra {\it et al.}~(\cite{mm7}, \cite{mmd2}, \cite{mmd22}),
and the references there in. Some significant works on general standby redundancy have also been developed by the researchers, namely, Cha {\it et al.}~\cite{cmy7}, Yun and Cha~\cite{yc7}, Li {\it et al.}~(\cite{lzw71}, \cite{lwz72}),
and Eryilmaz~\cite{es7}.
\\\hspace*{0.3 in}For an absolutely continuous random variable $X$, let the probability density function be denoted
 by $f_X(\cdot)$, the distribution function by $F_X(\cdot)$, the hazard rate function by $r_X(\cdot)$, 
and the reversed hazard rate function by $\tilde r_X(\cdot)$. We write $\bar F_X(\cdot)\equiv 1-F_X(\cdot)$ to denote the 
survival (or reliability) function of the random variable $X$. 
 \\\hspace*{0.3 in}Among all, Cha {et al.}~\cite{cmy7}, to the best of our knowledge, are the first to develop a new technique to handle the general standby redundancy based on the concept of accelerated life model (see, Nelson~\cite{n7})
 and virtual age model (see, Kijima~\cite{k71}, and Finkelstein~(\cite{f71}, \cite{f72})). Let $X$ be a random variable representing the lifetime 
 of a component, and $Y$ be another random variable representing the lifetime of a general standby redundancy in a usual environment. Assume that $X$ and $Y$ are independent.
 Consider the system where $Y$ is allocated to $X$. This system is called general standby system, and it is denoted by $X\cd Y$.   
 Further, let $Y^*$ be the lifetime of the redundancy in the milder environment. It is obvious that the lifetime of the redundancy in the milder environment is 
 stochastically larger than that in the usual environment. Thus, based on the idea used in the accelerated life model, 
 it is reasonable to assume that $F_{Y^*}(\cdot)=F_{Y}(\ga(\cdot))$, where $\ga(\cdot)$ satisfies:
 $(i)$ $0\leq \ga(t)\leq t$ for all $t\geq 0$, and $(ii)$ $\ga(t)$ is increasing in $t$.
 Further, suppose that the redundant component works in the milder environment 
 during the time $(0,t]$ without failure, and it gets activated in the usual environment at time $t$. Then, according to the virtual age model, the
 redundant component should have virtual age $\om(t)$, where $\om(\cdot)$ satisfies:
 $(i)$ $0\leq \om(t)\leq t$ for all $t\geq 0$,
 and $(ii)$ $\om(t)$ is increasing in $t$. It is worth to mention here that $\om(\cdot)$ should depend on $\ga(\cdot)$.
 To know more details about this, we refer the reader to see Cha {et al.}~\cite{cmy7}, Yun {\it et al.}~\cite{yc7}, and Li {\it et al.}~\cite{lwz72}.
 We call $\mathscr{A_{\ga,\om}}=\{\ga(\cdot)$, $\om(\cdot)\}$ as the {\it model function} associated with the system $X\cd Y$.
 According to Li {\it et al.}~\cite{lzw71}, the system $X\cd Y$ can be represented as
 \begin{eqnarray*}
  X\cd Y=
\left\{
	\begin{array}{ll}
		X  & \mbox{if } X\geq Y^* \\
		X+Y_{\om(X)} & \mbox{if } X<Y^*,
	\end{array}
\right.
 \end{eqnarray*}
 where $Y_{\om(X)}=\left[Y-\om(X)|Y>\om(X)\right]$ is the residual lifetime at random time (see, Cai and Zheng~\cite{cz7}, and the references there in).
 By Theorem 1 of Cha {et al.}~\cite{cmy7}, the reliability function is given by
 \begin{eqnarray}
  \bar F_{X\cd Y}(t)=\bar F_{X}(t)+\int\limits_0^t\frac{\bar F_{Y}(t-\delta(u))}{\bar F_{Y}(\omega(u))}\bar F_{Y}(\gamma(u))dF_{X}(u),\label{meq7}
 \end{eqnarray}
 where $\delta(u)=u-\omega(u)$ for all $u\geq 0$.
 It is worth to mention here that the general standby system reduces to the cold standby system when $\ga(t)=\om(t)\equiv0$, and 
  to the hot standby when $\ga(t)=\om(t)\equiv t$.
  \\\hspace*{0.3 in}Let us discuss a standby system which is more generalized than what we have discussed above. 
  Suppose, if possible, that we have some prior knowledge (which may be gained through experience or otherwise) about the original component that it will surely function
  at least for $t_0$ ($\geq 0$) units of time.
  In that case the redundant component should be activated in the milder environment at time $t_0$. Then the earlier system coincides with this
  system, only when we take $X_{t_0}$ in place of $X$ where $X_{t_0}=\left[X-t_0|X>t_0\right]$.
Furthermore, most of the results (what we will discuss in the 
   upcoming sections) also follow similarly for this system. 
 \\\hspace*{0.3 in}The reader may wonder about the fact $-$ how to calculate $\ga(\cdot)$ and $\om(\cdot)$ in a real situation! The $\ga(\cdot)$ function determines the environment under which the redundant component
 works at the initial stage. Thus, it is up to the user to decide which (milder) environment he/she wants to put the redundant component in, and $\ga(\cdot)$ will be fixed accordingly. 
 On the other hand, there must be some kind of indicating device (which have to be attached with the system) which will tell the user the time point when the redundant component 
 is being switched over to the usual environment from the milder environment.
 Consequently, $\om(\cdot)$ will be determined. 
 \\\hspace*{0.3 in}The stochastic orders are basically used to compare the lifetimes of two systems. In the literature many different types of stochastic 
 orders have been defined. Each stochastic order has its individual importance. The following well known definitions may be obtained in Shaked and Shanthikumar~\cite {shak1}.
\begin{definition}
Let $X$ and $Y$ be two absolutely continuous random variables with respective supports $(l_X,u_X)$ and $(l_Y,u_Y)$,
where $u_X$ and $u_Y$ may be positive infinity, and $l_X$ and $l_Y$ may be negative infinity.
Then, $X$ is said to be smaller than $Y$ in
\begin{enumerate}
\item likelihood ratio (lr) order, denoted as $X\leq_{lr}Y$, if
$$\frac{f_Y(t)}{f_X(t)}\;\rm{is\, increasing \, in} \,t\in(l_X,u_X)\cup(l_Y,u_Y);$$
\item hazard rate (hr) order, denoted as $X\leq_{hr}Y$, if $$\frac{\bar F_Y(t)}{\bar F_X(t)}\;\rm{is\,increasing \,in}\, t \in (-\infty,max(u_X,u_Y)),$$
\item reversed hazard rate (rhr) order, denoted as $X\leq_{rhr}Y$, if $$ \frac{F_Y(t)}{ F_X(t)}\;\rm{is\,increasing \,in}\, t \in(min(l_X,l_Y),\infty),$$
 \item usual stochastic (st) order, denoted as $X\leq_{st}Y$, if $\bar F_X(t)\leq \bar F_Y(t)$ \\$\rm{for\; all}\;t\in~(-\infty,\infty)$;
 \item increasing concave (icv) order, denoted as $X\leq_{icv}Y$, if $E[\xi(X)]\leq E[\xi(Y)]$, for all increasing concave functions $\xi(\cdot)$.
\end{enumerate}
\end{definition} 
Below we give the chain of implications among the above discussed stochastic orders.
\\
\hspace*{1.4 in}$~~~~~~~X\leq_{hr}Y$
\\\hspace*{1.7 in}$~~~~~~\uparrow ~~~~~~~\searrow$
\\\hspace*{1.4 in} $~~~~~~X\leq_{lr}Y~~\rightarrow~~X\leq_{st}Y~~\rightarrow~~X\leq_{icv}Y.$

\hspace{1.7 in}$~~~~~~\downarrow$~~~~~~~$\nearrow$

\hspace{1.4 in}$~~~~~~~~X\leq_{rhr}Y$
\\Like stochastic orders, stochastic precedence order is also a very useful tool to compare the lifetimes of two systems. The detailed study of this order 
 may be found in Singh and Misra~\cite{sm7}, and Boland {\it et al.}~\cite{bsc1}. Usual stochastic order does not always imply stochastic 
precedence order. If the two random variables are dependent then there is no implication between usual stochastic order and stochastic precedence order.
  \begin{definition}
   Let $X$ and $Y$ be two continuous random variables. Then, $X$ is said to be greater than $Y$ in stochastic precedence (sp) order, denoted as $X\geq_{sp}Y$, if
   \begin{eqnarray*}
   P(X>Y)\geq P(Y>X).
   \end{eqnarray*}
 \end{definition}
 \hspace*{0.3 in}The paper is organized as follows. In Section 2, we discuss three different models, namely, Model I, Model II and Model III. The first two models are constructed based on the allocation strategy of single general 
 standby component whereas the third model is based on two general standby components.
 In Section 3, we discuss series systems (which are produced through different allocation strategies of standby components) corresponding to Model I and Model II.
 We compare them with respect to the usual stochastic 
 and the stochastic precedence orders. We show that the best strategy to get the optimal series system is the allocation of stochastically strongest standby to the stochastically
 weakest one. In Section 4, we compare parallel systems (corresponding to Model~I, Model~II and Model~III) with respect to the usual stochastic order. We show that allocation of stochastically strongest standby to the stochastically
 strongest component is the best strategy in order to get the optimal parallel system.
\\\hspace*{0.3 in}Throughout the paper, increasing and decreasing properties are not used in strict sense.
For any differentiable function $k(\cdot)$, we write $k'(t)$ to denote the first derivative of $k(t)$ with respect to $t$.
The random variables considered in this paper are all nonnegative.
\section{Construction of Models}
In this section we discuss three models.
\\{\bf Model I:}
Let $X_1,X_2,\dots,X_n$ be the random variables representing the lifetimes of $n$ components forming a series (resp. parallel) system, and $Y$ be the 
random variable representing the lifetime of a general standby redundancy. Assume that $X_1,X_2,\dots,X_n$ and $Y$ are independent.
Suppose that we are interested to allocate $Y$ to the component of the series (resp. parallel) system so that the optimal (in sense of reliability) series (resp. parallel) system will come out.
Thus one can allocate $Y$ to any one of $X_1,X_2,\dots,X_n$. If we do that, then we have $n$ consecutive series (resp. parallel) systems by allocating $Y$ to
$X_1,X_2,\dots,X_n$, respectively. 
We define, for $i=1,2,\dots,n$, 
\begin{eqnarray*}
 &&{U}^{s}_{i}=\min\{X_1,\dots,X_{i-1},X_i\cd Y,X_{i+1},\dots,X_n\}
 \\&&{U}^p_{i}=\max\{X_1, \dots,X_{i-1},X_i\cd Y,X_{i+1},\dots,X_n\},
\end{eqnarray*}
where the operation $X_i\cd Y$ stands for general standby system as discussed in Section $1$.
Clearly, for each $i=1,2,\dots,n$,
${U}^{s}_{i}$ (resp. ${U}^p_{i}$) represents the lifetime of a series (resp. parallel) system with general standby redundancy allocated to the $i$th
component.\\
\\{\bf Model II:} Suppose that we have $n$ general standby redundancies $Y_1, Y_2,\dots,Y_n$ in place of single redundancy $Y$ (as considered in Model I). But, we
could use exactly one of them due to some constraints. 
If we allocate $Y_i$ (in place of $Y$) to $X_i$ in the same way as it is done in the above described model, then we have $n$ series (resp. parallel) systems which are defined as, for $i=1,2,\dots,n$,  
\begin{eqnarray*}
 &&V^{s}_{i}=\min\{X_1, \dots,X_{i-1},X_i \cd Y_i,X_{i+1},\dots,X_n\}
 \\&&V^p_{i}=\max\{X_1,\dots,X_{i-1},X_i\cd Y_i,X_{i+1},\dots,X_n\}.
\end{eqnarray*}
Clearly, if $Y_1,Y_2,\dots,$ $Y_n$ are identical, then
$\left\{V^{s}_{1},V^s_2,\dots,V^{s}_{n}\right\}$ $\left(\text{resp.}\;\left\{{V}^p_{1},V^p_2,\dots,V^p_n\right\}\right)$ 
reduces to
$\left\{U^s_{1},U^s_2,\dots,U^s_{n}\right\}$ $\left(\text{resp. }\left\{{U}^p_{1},U^p_2,\dots,{U}^p_{n}\right\}\right)$.\\
\\{\bf Model III:} Let us now consider a model which is slightly different from the ones previously discussed.
Consider a series (resp. parallel) system formed by $n$ components $X_1,X_2,\dots,X_n$. Suppose that we have two general standby redundancies $Y_1$ and $Y_2$ where
$X_1,X_2,\dots,$ $X_n,Y_1$ and $Y_2$ are independent.
Now, we are interested to allocate $Y_1$ and $Y_2$ with any two components of the system in such a way that system reliability becomes optimum. 
Without any loss of generality, let us allocate them with $X_1$ and $X_2$, and 
this could be done in the following two ways: either $X_1$ with $Y_2$ and $X_2$ with $Y_1$, or $X_1$ with $Y_1$ and $X_2$ with $Y_2$. 
If we do that, then we have two series (resp. parallel) systems $Q^s_1$ and $Q^s_2$ (resp. $Q^p_1$ and $Q^p_2$) 
which are defined as
\begin{eqnarray*}
 &&Q^s_1=\min\{X_1\cd Y_2, X_2\cd Y_1,X_3,\dots,X_n\}
 \\&&Q^s_2=\min\{X_1\cd Y_1,X_2\cd Y_2,X_3,\dots,X_n\}
\end{eqnarray*}
and
\begin{eqnarray*}
 &&Q^p_1=\max\{X_1\cd Y_2, X_2\cd Y_1,X_3,\dots,X_n\}
 \\&&Q^p_2=\max\{X_1\cd Y_1,X_2\cd Y_2,X_3,\dots,X_n\}.
\end{eqnarray*}
Many applications of the above three discussed models are found in reality. In case of hot standby and cold standby, many different researchers
have studied the above three models; see, for example, Boland {\it et al.}~\cite{bep7}, Singh and Misra~\cite{sm7}, Vald\'es and Zequeira~(\cite{vz71}, \cite{vz72}), Romera {\it et al.}~\cite{rvz2},
Li and Hu~\cite{lh22}, Brito {\it et al.}~\cite{br1}, Li {\it et al.}~\cite{lyh2}, Misra {\it et al.}~(\cite{mm7}, \cite{mmd2}, \cite{mmd22}) and the references there in.  
\\\hspace*{0.3 in}For the sake of mathematical simplicity, we write $H_1=\min\{X_3,X_4,\dots,X_n\}$ and $H_2=\max\{X_3,X_4,\dots,X_n\}$.
Further, for $i=1,2,\dots,n$, we write $\delta_i(u)=u-\omega_i(u)$ for all $u\geq0$.
\section{Series System} 
\hspace*{0.3 in}Both for hot and cold standby systems, Vald\'{e}s, J.E. and Zequeira~(\cite{vz71}, \cite{vz72}), Misra {\it et al.}~(\cite{mmd2}, \cite{mmd22})
and many other researchers have shown that allocation of stochastically strongest standby to the stochastically weakest component is the best strategy to get the optimal series system.
The following theorem shows that the same strategy also follows for general standby system. 
\begin{thm}\label{thw71}
Let $\mathscr{A}_{\ga_i,\om_i}$ be the model function
associated with $X_i\cd Y_i$, for $i=1,2$. Assume that $X_1\leq_{hr}X_2$ and $Y_2\leq_{hr}Y_1$. Suppose that one of the following conditions holds:
\begin{enumerate}
 \item [$(i)$] $\gamma_1(u)\leq\gamma_2(u)$ and $\om_1(u)=\om_2(u)$ for all $u\geq 0$.
 \item [$(ii)$] $Y_1$ or $Y_2$ have log-concave (log-convex) survival functions, 
 and $\omega_1(u)\leq\omega_2(u)$ ($\omega_1(u)\geq\omega_2(u)$) and $\gamma_1(u)\leq\gamma_2(u)$ for all $u\geq 0$.
 \end{enumerate}
 Then, $V^s_2$ $\leq_{st}~V^s_1$.
\end{thm}
{\bf Proof:} We first prove the result under $(i)$. Note that
\begin{eqnarray*}
 \bar F_{V^s_1}(t)=\bar F_{X_2}(t)\bar F_{H_1}(t)\bar F_{X_1\cd Y_1}(t)
\end{eqnarray*}
and 
\begin{eqnarray*}
 \bar F_{V^s_2}(t)=\bar F_{X_1}(t)\bar F_{H_1}(t)\bar F_{X_2\cd Y_2}(t).
\end{eqnarray*}
Writing $\Delta_1(t)=\bar F_{V^s_1}(t)-\bar F_{V^s_2}(t)$, we have, by (\ref{meq7}),
\begin{eqnarray}
 \Delta_1(t)&=&\bar F_{X_2}(t)\bar F_{H_1}(t)\int\limits_0^t\frac{\bar F_{Y_1}(t-\delta_1(u))}{\bar F_{Y_1}(\omega_1(u))}\bar F_{Y_1}(\gamma_1(u))dF_{X_1}(u)\nonumber
\\ &&-\bar F_{X_1}(t)\bar F_{H_1}(t)\int\limits_0^t\frac{\bar F_{Y_2}(t-\delta_2(u))}{\bar F_{Y_2}(\omega_2(u))}\bar F_{Y_2}(\gamma_2(u))dF_{X_2}(u).\label{ew71}
\end{eqnarray}
Since, $Y_2\leq_{hr}Y_1$ we have, for $0\leq u \leq t$,
\begin{eqnarray*}
\frac{\bar F_{Y_1}(t-\delta_1(u))}{\bar F_{Y_1}(\omega_1(u))}&\geq& \frac{\bar F_{Y_2}(t-\delta_1(u))}{\bar F_{Y_2}(\omega_1(u))}
\\&=&\frac{\bar F_{Y_2}(t-\delta_2(u))}{\bar F_{Y_2}(\omega_2(u))},
\end{eqnarray*}
where the equality follows from the fact that $\om_1(u)=\om_2(u)$ for all $u\geq0$. This gives
\begin{eqnarray}
\frac{\bar F_{Y_1}(t-\delta_1(u))}{\bar F_{Y_1}(\omega_1(u))}\bar F_{Y_1}(\gamma_1(u))\geq \frac{\bar F_{Y_2}(t-\delta_2(u))}{\bar F_{Y_2}(\omega_2(u))}\bar F_{Y_2}(\gamma_2(u)),\label{ew72}
\end{eqnarray}
because $\bar F_{Y_1}(u)\geq \bar F_{Y_2}(u)$ and $\gamma_1(u)\leq\gamma_2(u)$ for all $u\geq 0$. Thus, on using (\ref{ew72}) in (\ref{ew71}) we have
\begin{eqnarray}
 \Delta_1(t)&\geq&\bar F_{X_2}(t)\bar F_{H_1}(t)\int\limits_0^t\frac{\bar F_{Y_2}(t-\delta_2(u))}{\bar F_{Y_2}(\omega_2(u))}\bar F_{Y_2}(\gamma_2(u))dF_{X_1}(u)\nonumber 
 \\&&-\bar F_{X_1}(t)\bar F_{H_1}(t)\int\limits_0^t\frac{\bar F_{Y_2}(t-\delta_2(u))}{\bar F_{Y_2}(\omega_2(u))}\bar F_{Y_2}(\gamma_2(u))dF_{X_2}(u)\nonumber
 \\&=&\bar F_{H_1}(t)\int\limits_0^t\frac{\bar F_{Y_2}(t-\delta_2(u))}{\bar F_{Y_2}(\omega_2(u))}\bar F_{Y_2}(\gamma_2(u))\xi_t(u)du,\nonumber
\end{eqnarray}
where $\xi_t(u)=\bar F_{X_2}(t)f_{X_1}(u)-\bar F_{X_1}(t)f_{X_2}(u),\;0\leq u\leq t$. To prove $\Delta_1(t)\geq 0$, it suffices to show that
$\xi_t(u)\geq0$ for $0\leq u\leq t$.
Since, $X_1\leq_{hr}X_2$, it implies that
\begin{eqnarray*}
 r_{X_1}(u)\frac{\bar F_{X_2}(t)}{\bar F_{X_2}(u)}\geq r_{X_2}(u)\frac{\bar F_{X_1}(t)}{\bar F_{X_1}(u)},
\end{eqnarray*}
or equivalently,
\begin{eqnarray}
\xi_t(u)\geq0.\label{ew82}
\end{eqnarray}
Thus, $V^s_2\leq_{st}V^s_1$. To prove the result under $(ii)$ we proceed as follows. Consider the following two cases.
\\{ Case I:} Let $Y_1$ have log-concave (log-convex) survival function.
Then, for $0\leq u \leq t$ and $\omega_1(u)\leq\omega_2(u)$ ($\omega_1(u)\geq\omega_2(u)$),
\begin{eqnarray}
\frac{\bar F_{Y_1}(t-\delta_1(u))}{\bar F_{Y_1}(\omega_1(u))}&\geq& \frac{\bar F_{Y_1}(t-\delta_2(u))}{\bar F_{Y_1}(\omega_2(u))}\nonumber
\\&\geq&\frac{\bar F_{Y_2}(t-\delta_2(u))}{\bar F_{Y_2}(\omega_2(u))},\label{lw81}
\end{eqnarray}
where the first inequality holds because $Y_1$ has log-concave (log-convex) survival function. The second inequality follows from $Y_2\leq_{hr}Y_1$.
\\{ Case II:} Let $Y_2$ have log-concave (log-convex) survival function. Then, for $0\leq u \leq t$ and $\omega_1(u)\leq\omega_2(u)$ ($\omega_1(u)\geq\omega_2(u)$) we have
\begin{eqnarray}
\frac{\bar F_{Y_1}(t-\delta_1(u))}{\bar F_{Y_1}(\omega_1(u))}&\geq&\frac{\bar F_{Y_2}(t-\delta_1(u))}{\bar F_{Y_2}(\omega_1(u))}\nonumber
\\&\geq&\frac{\bar F_{Y_2}(t-\delta_2(u))}{\bar F_{Y_2}(\omega_2(u))},\label{lw83}
\end{eqnarray}
where the first inequality follows from $Y_2\leq_{hr}Y_1$, and the second inequality holds because $Y_2$ has log-concave (log-convex) survival function.
Again, $\bar F_{Y_1}(u)\geq \bar F_{Y_2}(u)$ and $\gamma_1(u)\leq\gamma_2(u)$ for all $u\geq 0$. Thus, by (\ref{lw81}) or (\ref{lw83}) we have (\ref{ew72}). Then we proceed exactly
in the same way as it is done in $(i)$, and we get $\Delta_1(t)\geq 0$. Hence
$V^s_2\leq_{st}~V^s_1$.$\hfill \Box$
\\\hspace*{0.3 in}Below we give an example which supports the above theorem.
\begin{example}
 Let $X_i$, $i=1,2,\dots,n$, be the independent component lives having reliabilities $\bar F_{X_i}(t)=e^{-\lambda_i t}$, $t>0$, $\lambda_i>0$,
 with $\lambda_1\geq \lambda_2$, and $Y_j$, $j=1,2$ be the independent lives of the redundancies having reliabilities
 $\bar F_{Y_j}(t)=e^{-\mu_j t}$, $t>0$ with $0<\mu_1\leq \mu_2$. Further, let $X_i$'s and $Y_j$'s be independent. Clearly, $X_1\leq_{hr}X_2$ and $Y_2\leq_{hr}Y_1$.
 Assume that, for $j=1,2$, and for all $u\geq 0$, $\ga_j(u)=a_ju$
 and $\omega_j(u)=b_ju$, where $0< a_1\leq a_2\leq b_1\leq b_2\leq 1$.
 It is easy to verify that
all the conditions given in Theorem~\ref{thw71} are satisfied. Hence $V^s_{2}\leq_{st}V^s_{1}$.$\hfill \Box$
\end{example}
\hspace*{0.3 in}The following theorem is a natural extension of the above theorem. 
\begin{thm}
Let $\mathscr{A}_{\ga_i,\om_i}$ be the model function 
associated with $X_i\cd Y_i$, for $i=1,2,\dots,n$.
Assume that $X_1\leq_{hr}X_2\leq_{hr}\dots\leq_{hr}X_n$ and $Y_n\leq_{hr}\dots \leq_{hr}Y_2\leq_{hr}Y_1$. Suppose that one of the following conditions holds:
\begin{enumerate}
 \item [$(i)$] $\gamma_1(u)\leq\gamma_2(u)\leq\dots\leq\ga_n(u)$ and $\om_1(u)=\om_2(u)=\dots=\om_n(u)$ for all $u\geq 0$.
 \item [$(ii)$] Let $n$ be an even integer. Further, $Y_2,Y_4,\dots,Y_{n}$ or $Y_1,Y_3,\dots,Y_{n-1}$ have log-concave (log-convex) survival functions,
 and $\omega_1(u)\leq\omega_2(u)\leq\dots\leq\om_n(u)$ ($\omega_1(u)\geq\omega_2(u)\geq\dots\geq\om_n(u)$) and $\gamma_1(u)\leq\gamma_2(u)\leq\dots\leq\ga_n(u)$ for all $u\geq 0$.
 \item [$(iii)$] Let $n$ be an odd integer. Further, $Y_2,Y_4,\dots,Y_{n-1}$ or $Y_1,Y_3,\dots,Y_{n}$ have log-concave (log-convex) survival functions,
 and $\omega_1(u)\leq\omega_2(u)\leq\dots\leq\om_n(u)$ ($\omega_1(u)\geq\omega_2(u)\geq\dots\geq\om_n(u)$) and $\gamma_1(u)\leq\gamma_2(u)\leq\dots\leq\ga_n(u)$ for all $u\geq 0$.
 \end{enumerate}
 Then, $V^s_n\leq_{st}\dots\leq_{st}V^s_2\leq_{st}V^s_1$.$\hfill \Box$
\end{thm}
\hspace*{0.3 in}The following corollary is an immediate consequence of the above theorem.
\begin{coro}
Let $\mathscr{A}_{\ga_i,\om_i}$ be the model function 
associated with $X_i\cd Y$, for $i=1,2,\dots,n$.
Assume that $X_1\leq_{hr}X_2\leq_{hr}\dots\leq_{hr}X_n$. Suppose that one of the following conditions holds:
\begin{enumerate}
 \item [$(i)$] $\gamma_1(u)\leq\gamma_2(u)\leq\dots\leq\ga_n(u)$ and $\om_1(u)=\om_2(u)=\dots=\om_n(u)$ for all $u\geq 0$.
 \item [$(ii)$] $Y$ has log-concave (log-convex) survival function, and $\omega_1(u)\leq\omega_2(u)\leq\dots\leq\om_n(u)$ ($\omega_1(u)\geq\omega_2(u)\geq\dots\geq\om_n(u)$)
 and $\gamma_1(u)\leq\gamma_2(u)\leq\dots\leq\ga_n(u)$ for all $u\geq 0$.
 \end{enumerate}
 Then, $U^s_{n}\leq_{st}\dots\leq_{st}U^s_{2}\leq_{st}U^s_{1}$.$\hfill \Box$
\end{coro}
\hspace*{0.3 in}In the following theorem we show that the condition $Y_2\leq_{hr} Y_1$ given in Theorem~\ref{thw71} can be replaced by $Y_2\leq_{st}Y_1$ under
some additional restriction on the model
function.
\begin{thm}
Let $X_1\cd Y_1$ and $X_2\cd Y_2$ have the same model function $\mathscr{A}_{\ga,\om}$.
Assume that $\delta (u)$ is increasing and $\omega(u)= \gamma(u)$ for $u\geq 0.$ If $X_1\leq_{hr}X_2$ and
 $Y_2\leq_{st}Y_1$ then $V^s_2\leq_{st}V^s_1$.
\end{thm}
{\bf Proof:} 
 From (\ref{ew71}) we have
\begin{eqnarray*}
 \Delta_1(t)&=&\bar F_{X_2}(t)\bar F_{H_1}(t)\int\limits_0^t\bar F_{Y_1}(t-\delta(u))dF_{X_1}(u)
\\ &&-\bar F_{X_1}(t)\bar F_{H_1}(t)\int\limits_0^t \bar F_{Y_2}(t-\delta(u))dF_{X_2}(u)
\\&\geq&\bar F_{H_1}(t)\int\limits_0^t\bar F_{Y_2}(t-\delta(u))\xi_t(u)du
\\&\geq &0,
\end{eqnarray*}
 where
the first inequality follows from $Y_2\leq_{st}Y_1$ and the second inequality follows from (\ref{ew82}). Thus, $V^s_2\leq_{st}V^s_1$.$\hfill \Box$
\\\hspace*{0.3 in}The following generalization of the above theorem is obvious.
\begin{coro}
Let each of $X_1\cd Y_1,X_2\cd Y_2,\dots,X_n\cd Y_n$ have the same model function $\mathscr{A}_{\ga,\om}$.
 Assume that $\delta (u)$ is increasing and $\omega(u)= \gamma(u)$ for $u\geq 0$. If $X_1\leq_{hr}X_2\leq_{hr}\dots \leq_{hr}X_n$ and
 $Y_n\leq_{st}\dots \leq_{st}Y_2 \leq_{st}Y_1$ then $V^s_n\leq_{st}\dots \leq_{st}V^s_2\leq_{st}V^s_1$.$\hfill \Box$
\end{coro}
\hspace*{0.3 in}As a consequence of the above theorem we have the following corollary.
\begin{coro}
Let each of $X_1\cd Y,X_2\cd Y,\dots,X_n\cd Y$ have the same model function $\mathscr{A}_{\ga,\om}$.
 Assume that $\delta (u)$ is increasing and $\omega(u)= \gamma(u)$ for $u\geq 0$. 
 If $X_1\leq_{hr}X_2\leq_{hr}\dots \leq_{hr}X_n$ then $U^s_{n}\leq_{st}\dots \leq_{st}U^s_{2}\leq_{st}U^s_{1}$.$\hfill \Box$
\end{coro}
\begin{remark}
 Let us discuss one such situation where $\ga(t)=\om(t)$ holds. Consider the system $X\cd Y$ as discussed in Section 1. Assume that the lifetime of $Y$ in a milder environment during the time $[0,t]$
 is same as that of $Y$ in the usual environment during the time $[0,\om(t)]$. Then, for all $t\geq 0$
 $F_{Y^*}(t)=F_Y(\om(t))$, or equivalently, $\ga(t)=\om(t)$
 (cf. Nelson~\cite{n7}, and Yun and Cha~\cite{yc7}). $\hfill \Box$
\end{remark}
\hspace*{0.3 in}As we know, there is no relation between the usual stochastic order and the stochastic precedence order for dependent systems. Thus, the natural question arises $-$ 
whether the above result holds for the stochastic precedence order. The following theorem gives an affirmative answer of this question. Before stating the theorem
 we give two lemmas which will be used in proving the theorem. The first lemma is taken from Barlow and Proschan~(\cite{bp1}, p. 120).
The proof of the second lemma follows from definition of the increasing concave order.
\begin{lem}\label{le73}
 Let $W(u)$ be a Lebesgue-Stieltjes measure, not necessarily positive, for which $\int\limits_{0}^t dW(u)\geq 0$ for all $t$, and let $\eta(u)$ be nonnegative and decreasing in $u\geq 0$. Then 
 $\int\limits_{0}^{\infty}\eta(u)dW(u)\geq 0.$ $\hfill \Box$
\end{lem}
\begin{lem}\label{ledcn}
  Let $X$ and $Y$ be independent random variables. Then $X\leq_{icv}Y$ if, and only if,  $E\left(g(X)\right)\geq E\left(g(Y)\right)$ for all decreasing convex function $g(\cdot)$.$\hfill \Box$
\end{lem}
\begin{thm}\label{ut770}
Let $\mathscr{A}_{\ga_i,\om_i}$ be the model function 
associated with $X_i\cd Y_i$, for $i=1,2$.
  Suppose that one of the following conditions holds:
 \begin{enumerate}
 \item [$(i)$] Let $\gamma_1(u)\leq\gamma_2(u)$ for all $u\geq0$, and $X_1\leq_{st}X_2$ and $Y_2\leq_{st}Y_1$.
 \item [$(ii)$] $X_1\leq_{icv}X_2$ and $Y^*_2\leq_{icv}Y^*_1$, and $Y^*_2,X_1,X_3,\dots,X_n$ (or $Y^*_1,X_2,X_3,\dots,X_n$) have convex survival functions.
\end{enumerate}
Then $V^s_2\leq_{sp}V^s_1$.
\end{thm}
{\bf Proof:} We first prove the result under condition $(i)$. Note that
\begin{eqnarray}
 P\left(V^s_1>V^s_2\right)&=&P\left(\min\{X_1\cd Y_1,X_2,H_1\}>\min\{X_1,X_2\cd Y_2,H_1\}\right)\nonumber
 \\&=&P\left(Y^*_1>X_1, X_2>X_1,H_1>X_1\right).\label{ews18}
\end{eqnarray}
Writing $\Delta_2=P\left(V^s_{1}>V^s_{2}\right)-P\left(V^s_{2}>V^s_{1}\right)$ we have
\begin{eqnarray}
 \Delta_2&=&\left(Y^*_1>X_1, X_2>X_1,H_1>X_1\right)-\left(Y^*_2>X_2, X_1>X_2,H_1>X_2\right)\nonumber
 \\&=&\int\limits_0^{\infty}\bar F_{Y_1}(\ga_1(u))\bar F_{X_2}(u)\bar F_{H_1}(u)dF_{X_1}(u)\nonumber
 \\&&-\int\limits_0^{\infty}\bar F_{Y_2}(\ga_2(u))\bar F_{X_1}(u)\bar F_{H_1}(u)dF_{X_2}(u)\label{ews21}
 \\&\geq&\int\limits_0^{\infty}\bar F_{Y_2}(\ga_2(u))\bar F_{X_1}(u)\bar F_{H_1}(u)d[F_{X_1}(u)-F_{X_2}(u)],\label{ews19}
\end{eqnarray}
where the inequality follows from the fact that $\bar F_{Y_1}(\ga_1(u))\geq \bar F_{Y_2}(\ga_2(u))$ and $\bar F_{X_2}(u)\geq \bar F_{X_1}(u)$ for all $u\geq 0$. Further, 
$X_1\leq_{st}X_2$ implies that, for any fixed $t\geq 0$,
\begin{eqnarray}
 \int\limits_0^{t}d[F_{X_1}(u)-F_{X_2}(u)]\geq 0.\label{ews20}
\end{eqnarray}
Since, $\bar F_{Y_2}(\ga_2(u))\bar F_{X_1}(u)\bar F_{H_1}(u)$ is nonnegative and decreasing in $u\geq 0$,
then, on using (\ref{ews20}) in (\ref{ews19}) we have, from Lemma~\ref{le73}, $\Delta_2\geq 0$. To prove the result under condition $(ii)$, we proceed as follows. 
We consider following two cases.
\\{Case I:} $Y^*_2,X_1,X_3,\dots,X_n$ have convex survival functions.
Note that (\ref{ews21}) can equivalently be written as
$$\Delta_2=\Im_1+\Im_2,$$
where 
\begin{eqnarray}
 \Im_1=\int\limits_0^{\infty}\left[\bar F_{Y_1}(\ga_1(u))\bar F_{X_2}(u)-\bar F_{Y_2}(\ga_2(u))\bar F_{X_1}(u)\right]\bar F_{H_1}(u)f_{X_1}(u)du\label{ews22}
\end{eqnarray}
and
\begin{eqnarray*}
 \Im_2&=&\int\limits_0^{\infty}\bar F_{Y_2}(\ga_2(u))\bar F_{X_1}(u)\bar F_{H_1}(u)\left[dF_{X_1}(u)-dF_{X_2}(u)\right]\nonumber
 \\&=&E\left[\bar F_{Y^*_2}(X_1)\bar F_{X_1}(X_1)\bar F_{H_1}(X_1)\right]-E\left[\bar F_{Y^*_2}(X_2)\bar F_{X_1}(X_2)\bar F_{H_1}(X_2)\right]\label{ews23}
\end{eqnarray*}
Since, increasing concave order is closed under minimum (see, Corollary 4.A.16 of Shaked and Shanthikumar~\cite{shak1}), by $X_1\leq_{icv}X_2$ and $Y^*_2\leq_{icv}Y^*_1$ we have, for all $t\geq0$,
\begin{eqnarray}
 \int\limits_0^{t}\left[\bar F_{Y_1}(\ga_1(u))\bar F_{X_2}(u)-\bar F_{Y_2}(\ga_2(u))\bar F_{X_1}(u)\right]du\geq 0.\label{ews24}
\end{eqnarray}
Again, by hypotheses, $\bar F_{H_1}(u)f_{X_1}(u)$ is nonnegative and decreasing in $u\geq 0$. So, on using (\ref{ews24}) in (\ref{ews22}) we have, from Lemma~\ref{le73}, $\Im_1\geq 0$.
Further, by hypotheses, $\bar F_{Y^*_2}(u)\bar F_{X_1}(u)\bar F_{H_1}(u)$ is decreasing and convex. Then, by Lemma~\ref{ledcn}, we have $\Im_2\geq0$, because $X_1\leq_{icv}X_2$. 
Thus, $\Delta_2\geq 0$,
and hence $V^s_{2}\leq_{sp}V^s_{1}$.
\\Case II:  $Y^*_1,X_2,X_3,\dots,X_n$ have convex survival functions. Now, (\ref{ews21}) could be written as
$$\Delta_2=\Im_3+\Im_4,$$
where 
\begin{eqnarray*}
 \Im_3=\int\limits_0^{\infty}\left[\bar F_{Y_1}(\ga_1(u))\bar F_{X_2}(u)-\bar F_{Y_2}(\ga_2(u))\bar F_{X_1}(u)\right]\bar F_{H_1}(u)f_{X_2}(u)du\label{ews25}
\end{eqnarray*}
and
\begin{eqnarray*}
 \Im_4&=&\int\limits_0^{\infty}\bar F_{Y_1}(\ga_1(u))\bar F_{X_2}(u)\bar F_{H_1}(u)\left[dF_{X_1}(u)-dF_{X_2}(u)\right]\nonumber
 \\&=&E\left[\bar F_{Y^*_1}(X_1)\bar F_{X_2}(X_1)\bar F_{H_1}(X_1)\right]-E\left[\bar F_{Y^*_1}(X_2)\bar F_{X_2}(X_2)\bar F_{H_1}(X_2)\right].\label{ews26}
\end{eqnarray*}
Proceeding in the same way as in Case I, we get $\Im_3\geq 0$ and $\Im_4\geq0$, and hence $\Delta_2\geq 0$. Thus, $V^s_{2}\leq_{sp}V^s_{1}$.$\hfill \Box$ 
\\\hspace*{0.3 in}The following corollary is an immediate consequence of the above theorem.
\begin{coro}
Let $\mathscr{A}_{\ga_i,\om_i}$ be the model function 
associated with $X_i\cd Y$, for $i=1,2$.
  Suppose that one of the following conditions holds:
 \begin{enumerate}
 \item [$(i)$] $\gamma_1(u)\leq\gamma_2(u)$ for all $u\geq0$, and $X_1\leq_{st}X_2$.
 \item [$(ii)$] $\gamma_1(u)=\gamma_2(u)$ for all $u\geq0$. Further, $X_1\leq_{icv}X_2$ and $Y^*,X_1,X_3,\dots,X_n$ (or $Y^*,X_2,X_3,\dots,X_n$) have convex survival function.
\end{enumerate}
Then $U^s_2\leq_{sp}U^s_1$.$\hfill \Box$
\end{coro}
\hspace*{0.3 in}Below we give an example which supports the above theorem.  
\begin{example}
 Let $X_i$, $i=1,2,\dots,n$, be the independent component lives having reliabilities $\bar F_{X_i}(t)=e^{-\lambda_i t}$, $t>0$, $\lambda_i>0$, with $\lambda_1\geq \lambda_2$,
 and $Y_j$, $j=1,2$ be the independent lives of the redundancies having reliabilities
 $\bar F_{Y_j}(t)=e^{-\mu_j t}$, $t>0$ with $0<\mu_1\leq \mu_2$. Further, let $X_i$'s and $Y_j$'s be independent. Clearly, $X_1\leq_{st}X_2$. 
 Assume that, for $j=1,2$, and for all $u\geq 0$, $\ga_j(u)=a_j\log(1+u)$
 and $\omega_j(u)=b_j\log(1+u)$, where $0< a_1\leq a_2\leq b_1\leq b_2\leq 1$.
Then we have, for $j=1,2$, $\bar F_{Y_j^*}(t)=\bar F_{Y_j}(\ga_j(t))=1/(1+t)^{a_j\mu_j}$, which gives $Y^*_2\leq_{st}Y^*_1$. It is easy to verify that
all of $Y^*_j,X_i$ have the convex survival functions. Thus, all the conditions given in Theorem~\ref{ut770} are satisfied, and hence $V^s_{2}\leq_{sp}V^s_{1}$.$\hfill \Box$
\end{example}
\section{Parallel System}
\hspace*{0.3 in}In this section we discuss parallel system. We show that allocation of stochastically strongest standby to the stochastically strongest 
component is the best strategy for an optimal parallel system. Before going into details of the next theorem, we give the following lemma which may be obtained in Li \emph{et al.}~(\cite{lzw71}, proof of Theorem 2).
\begin{lem}\label{lws2}
 Let $\delta (u)$ and $\omega(u)- \gamma(u)$ be increasing in $u\geq 0$. Assume that $X$ has log-concave survival function. Then, for any fixed $t\geq 0$,
 $\frac{\bar F_{X}(t-\delta(u))}{\bar F_{X}(\omega(u))}\bar F_{X}(\gamma(u))$ is nonnegative and increasing in $u\in[0,t]$.$\hfill \Box$
\end{lem}
\hspace*{0.3 in}The following theorem shows that under some sufficient conditions, ${V}^p_{2}$ is greater than ${V}^p_{1}$ in usual stochastic order.
\begin{thm}\label{thw23}
Let $X_1\cd Y_1$ and $X_2\cd Y_2$ have the same model function $\mathscr{A}_{\ga,\om}$.
  Assume that $\delta (u)$ and $\omega(u)- \gamma(u)$ are increasing in $u\geq 0$.
 If $X_1\leq_{rhr}X_2$ and $Y_1\leq_{hr}Y_2$, and $Y_1$ or $Y_2$ have log-concave survival functions
 then ${V}^p_{1}\leq_{st}{V}^p_{2}$.
\end{thm}
{\bf Proof:} $Y_1\leq_{hr}Y_2$ implies that, for $0\leq u\leq~t$,
$$\frac{\bar F_{Y_2}(t-\delta(u))}{\bar F_{Y_2}(\omega(u))}\geq \frac{\bar F_{Y_1}(t-\delta(u))}{\bar F_{Y_1}(\omega(u))},$$
which gives
\begin{eqnarray}
\frac{\bar F_{Y_2}(t-\delta(u))}{\bar F_{Y_2}(\omega(u))}\bar F_{Y_2}(\gamma(u))\geq \frac{\bar F_{Y_1}(t-\delta(u))}{\bar F_{Y_1}(\omega(u))}\bar F_{Y_1}(\gamma(u)).\label{ews2}
\end{eqnarray}
 Let us consider the following two cases.
\\{ Case I:} Let $Y_1$ have log-concave survival function.
Writing $\Delta_3(t)=F_{{V}^p_1}(t)-F_{V^p_2}(t)$ we have
\begin{eqnarray}
 \Delta_3(t)&=&F_{X_2}(t) F_{H_2}(t)\left[ F_{X_1}(t)-\int\limits_0^t\frac{\bar F_{Y_1}(t-\delta(u))}{\bar F_{Y_1}(\omega(u))}\bar F_{Y_1}(\gamma(u))dF_{X_1}(u)\right]\nonumber
\\ &&-F_{X_1}(t)F_{H_2}(t)\left[F_{X_2}(t)-\int\limits_0^t\frac{\bar F_{Y_2}(t-\delta(u))}{\bar F_{Y_2}(\omega(u))}\bar F_{Y_2}(\gamma(u))dF_{X_2}(u)\right]\label{ews39}
\\&\geq &F_{X_2}(t) F_{H_2}(t)\left[ F_{X_1}(t)-\int\limits_0^t\frac{\bar F_{Y_1}(t-\delta(u))}{\bar F_{Y_1}(\omega(u))}\bar F_{Y_1}(\gamma(u))dF_{X_1}(u)\right]\nonumber
\\ &&-F_{X_1}(t)F_{H_2}(t)\left[F_{X_2}(t)-\int\limits_0^t\frac{\bar F_{Y_1}(t-\delta(u))}{\bar F_{Y_1}(\omega(u))}\bar F_{Y_1}(\gamma(u))dF_{X_2}(u)\right]\nonumber
\\&= & F_{H_2}(t)\int\limits_0^t\left[F_{X_2}(t)F_{X_1}(u)-F_{X_1}(t)F_{X_2}(u)\right]d\left(\frac{\bar F_{Y_1}(t-\delta(u))}{\bar F_{Y_1}(\omega(u))}\bar F_{Y_1}(\gamma(u))\right),\label{ews1}
\end{eqnarray}
where the inequality follows from (\ref{ews2}). Again, $X_1\leq_{rhr}X_2$ implies that, for $u\in[0,t]$,
\begin{eqnarray}
 F_{X_2}(t)F_{X_1}(u)-F_{X_1}(t)F_{X_2}(u)\geq 0.\label{ews12}
\end{eqnarray}
Hence, Lemma \ref{lws2} along with (\ref{ews12}), gives that
$\Delta_3(t)\geq 0,$ and hence ${V}^p_{1}\leq_{st}{V}^p_{2}$.
\\{Case II:} Let $Y_2$ have log-concave survival function. By (\ref{ews2}), $\Delta_3(t)$ can be written as
\begin{eqnarray*}
 \Delta_3(t)&\geq &F_{X_2}(t) F_{H_2}(t)\left[ F_{X_1}(t)-\int\limits_0^t\frac{\bar F_{Y_2}(t-\delta(u))}{\bar F_{Y_2}(\omega(u))}\bar F_{Y_2}(\gamma(u))dF_{X_1}(u)\right]\nonumber
\\ &&-F_{X_1}(t)F_{H_2}(t)\left[F_{X_2}(t)-\int\limits_0^t\frac{\bar F_{Y_2}(t-\delta(u))}{\bar F_{Y_2}(\omega(u))}\bar F_{Y_2}(\gamma(u))dF_{X_2}(u)\right]\nonumber
\\&= & F_{H_2}(t)\int\limits_0^t\left[F_{X_2}(t)F_{X_1}(u)-F_{X_1}(t)F_{X_2}(u)\right]d\left(\frac{\bar F_{Y_2}(t-\delta(u))}{\bar F_{Y_2}(\omega(u))}\bar F_{Y_2}(\gamma(u))\right)\label{ews38}
\end{eqnarray*}
Now, Lemma \ref{lws2} along with (\ref{ews12}), gives that
$\Delta_3(t)\geq 0,$ and hence ${V}^p_{1}\leq_{st}{V}^p_{2}$.$\hfill \Box$
\\\hspace*{0.3 in}Below we cite an example of the above theorem.
\begin{example}\label{pexs00}
 Let $X_i$, $i=1,2,\dots,n$, be the independent component lives having reliabilities $\bar F_{X_i}(t)=e^{-\lambda_i t}$, $t>0$, $\lambda_i>0$, with $\lambda_1\geq \lambda_2$, 
 and $Y_j$, $j=1,2$ be the independent lives of the redundancies having reliabilities
 $\bar F_{Y_j}(t)=e^{-\mu_j t}$, $t>0$, $\mu_j>0$, with $\mu_1\geq \mu_2$. Further, let $X_i$'s and $Y_j$'s be independent. Clearly, $X_1\leq_{rhr}X_2$ and $Y_1\leq_{hr}Y_2$,
 and all of $Y_1,Y_2$ have the log-concave survival functions.
 Assume that, for all $u\geq 0$, $\ga(u)=a\log(1+u)$
 and $\omega(u)=bu$, where $0< a\leq b\leq 1$.
Then, $\delta (u)$ and $\omega(u)- \gamma(u)$ are increasing in $u\geq 0$.
 Thus, all the conditions given in Theorem~\ref{thw23} are satisfied, and hence $V^p_{1}\leq_{st}V^p_{2}$.$\hfill \Box$
\end{example}
\hspace*{0.3 in}The following generalization of Theorem~\ref{thw23} is straightforward.
\begin{thm}\label{thwr23}
Let each of $X_1\cd Y_1,X_2\cd Y_2,\dots,X_n\cd Y_n$ have the same model function $\mathscr{A}_{\ga,\om}$.
 Assume that $\delta (u)$ and $\omega(u)-\gamma(u)$ are increasing in $u\geq 0$.
 Further, let $X_1\leq_{rhr}X_2\leq_{rhr}\dots \leq_{rhr}X_n$ and $Y_1\leq_{hr}Y_2\leq_{hr}\dots\leq_{hr}Y_n$.
 Suppose that one of the following conditions holds:
 \begin{enumerate}
 \item [$(i)$] Let $n$ be an even integer, and $Y_2,Y_4,\dots,Y_{n}$ or $Y_1,Y_3,\dots,Y_{n-1}$ have log-concave survival functions.
 \item [$(i)$] Let $n$ be an odd integer, and $Y_2,Y_4,\dots,Y_{n-1}$ or $Y_1,Y_3,\dots,Y_n$ have log-concave survival functions.
 \end{enumerate}
  Then, $V^p_1\leq_{st}V^p_2\leq_{st}\dots\leq_{st}V^p_n$.$\hfill \Box$
\end{thm}
\hspace*{0.3 in}The following corollary given in Li \emph{et al.}~\cite{lwz72} directly follows from Theorem~\ref{thwr23}.
\begin{coro}
  Let each of $X_1\cd Y,X_2\cd Y,\dots,X_n\cd Y$ have the same model function $\mathscr{A}_{\ga,\om}$.
 Assume that $\delta (u)$ and $\omega(u)-\gamma(u)$ are increasing in $u\geq 0$.
 If $X_1\leq_{rhr}X_2\leq_{rhr}\dots \leq_{rhr}X_n$ and $Y$ has log-concave survival function,
  then $U^p_1\leq_{st}U^p_2\leq_{st}\dots\leq_{st}U^p_n$.$\hfill \Box$
\end{coro}
\hspace*{0.3 in}The following theorem may be compared with Theorem~\ref{thw23}. Here we discuss the same result as in Theorem~\ref{thw23} under some weak condition.
\begin{thm}\label{thwr20}
Let $X_1\cd Y_1$ and $X_2\cd Y_2$ have the same model function $\mathscr{A}_{\ga,\om}$.
Assume that $\delta (u)$ is increasing and $\omega(u)= \gamma(u)$ for all $u\geq 0.$ If $X_1\leq_{rhr}X_2$ and
 $Y_1\leq_{st}Y_2$ then ${V}^p_{1}\leq_{st}{V}^p_{2}$.
\end{thm}
{\bf Proof:} From (\ref{ews39}) we have
\begin{eqnarray}
 \Delta_3(t)&=&F_{X_2}(t) F_{H_2}(t)\left[ F_{X_1}(t)-\int\limits_0^t\bar F_{Y_1}(t-\delta(u))dF_{X_1}(u)\right]\nonumber
\\ &&-F_{X_1}(t)F_{H_2}(t)\left[F_{X_2}(t)-\int\limits_0^t\bar F_{Y_2}(t-\delta(u))dF_{X_2}(u)\right]\nonumber
\\&\geq &F_{X_2}(t) F_{H_2}(t)\left[ F_{X_1}(t)-\int\limits_0^t\bar F_{Y_1}(t-\delta(u))dF_{X_1}(u)\right]\nonumber
\\ &&-F_{X_1}(t)F_{H_2}(t)\left[F_{X_2}(t)-\int\limits_0^t\bar F_{Y_1}(t-\delta(u))dF_{X_2}(u)\right]\nonumber
\\&= & F_{H_2}(t)\int\limits_0^t\left[F_{X_2}(t)F_{X_1}(u)-F_{X_1}(t)F_{X_2}(u)\right]d\bar F_{Y_1}(t-\delta(u)),\label{ews40}
\end{eqnarray}
where the inequality follows from $Y_1\leq_{st}Y_2$. Again, $\bar F_{Y_1}(t-\delta(u))$ is increasing in $u\in[0,t]$. Thus, $\Delta_3(t)\geq 0$ follows from (\ref{ews12}), 
and hence ${V}^p_{1}\leq_{st}{V}^p_{2}$.$\hfill \Box$
\\\hspace*{0.3 in}The following generalization of the above theorem is quite obvious.
\begin{thm}\label{thmdc}
Let each of $X_1\cd Y_1,X_2\cd Y_2,\dots,X_n\cd Y_n$ have the same model function $\mathscr{A}_{\ga,\om}$.
 Assume that $\delta (u)$ is increasing and $\omega(u)= \gamma(u)$ for $u\geq 0$. If $X_1\leq_{rhr}X_2\leq_{rhr}\dots \leq_{rhr}X_n$ and
 $Y_1\leq_{st}Y_2\leq_{st} \dots\leq_{st}Y_n$, then $V^p_1\leq_{st}V^p_2\leq_{st}\dots\leq_{st}V^p_n$.$\hfill \Box$
\end{thm}
\hspace*{0.3 in}As a corollary to Theorem~\ref{thmdc} we have the following result.
\begin{coro}
 Let each of $X_1\cd Y,X_2\cd Y,\dots,X_n\cd Y$ have the same model function $\mathscr{A}_{\ga,\om}$.
 Assume that $\delta (u)$ is increasing and $\omega(u)= \gamma(u)$ for $u\geq 0$. 
 If $X_1\leq_{rhr}X_2\leq_{rhr}\dots \leq_{rhr}X_n$ then $U^p_1\leq_{st}U^p_2\leq_{st}\dots\leq_{st}U^p_n$.$\hfill \Box$
\end{coro}
\hspace*{0.3 in}Whatever the results we have discussed so far, either they are associated with Model I or Model II.
In the following theorem we give a result which is associated with Model III.
Here, we discuss the systems of two standby components. 
For hot and cold standbys, these systems are widely studied by different researchers, namely, Boland {\it et al.}~\cite{bep7}, Singh and Misra~\cite{sm7}, Vald\'es and Zequeira~\cite{vz72},
Misra {\it et al.}~(\cite{mm7}, \cite{mmd2}, \cite{mmd22}) and the references there in.
Recently, Li {\it et al.}~\cite{lwz72} have showed that $Q^s_1\leq_{st}Q^s_2$ under the condition that $X_1\geq_{lr}X_2$ and $Y_1\geq_{hr}Y_2$. In consequence of this result we have the following
theorem where we compare parallel systems instead of series systems. 
\begin{thm}\label{thw67}
 Let each of $X_1\cd Y_1,X_2\cd Y_2,X_1\cd Y_2,X_2\cd Y_1$ have the same model function $\mathscr{A}_{\ga,\om}$.
 Assume that $\delta (u)$ is increasing and $\omega(u)= \gamma(u)$ for $u\geq 0$. Suppose that one of the following conditions holds:
 \begin{enumerate}
 \item [$(i)$] $X_1\leq_{lr}X_2$ and $Y_1\leq_{rhr}Y_2$.
  \item [$(ii)$] $X_1\geq_{lr}X_2$ and $Y_1\geq_{rhr}Y_2$.
 \end{enumerate}
Then, $Q^p_1\leq_{st}Q^p_2$.
\end{thm}
{\bf Proof:} Note that 
\begin{eqnarray*}
 F_{Q^p_1}(t)&=&F_{H_2}(t)\left[\int\limits_0^t F_{Y_2}(t-\delta(u))dF_{X_1}(u)\right]\left[\int\limits_0^t F_{Y_1}(t-\delta(u))dF_{X_2}(u)\right]
 \\&=&F_{H_2}(t)\int\limits_0^t\int\limits_0^t F_{Y_2}(t-\delta(u)) F_{Y_1}(t-\delta(v))f_{X_1}(u)f_{X_2}(v)dudv.
\end{eqnarray*}
Similarly,
\begin{eqnarray*}
 F_{Q^p_2}(t)
 &=&F_{H_2}(t)\int\limits_0^t\int\limits_0^t F_{Y_2}(t-\delta(u)) F_{Y_1}(t-\delta(v))f_{X_1}(v)f_{X_2}(u)dudv.
\end{eqnarray*}
Writing $\Delta_4(t)=F_{Q^p_1}(t)-F_{Q^p_2}(t)$ we have
\begin{eqnarray*}
 \Delta_4(t)=F_{H_2}(t)\int\limits_0^t\int\limits_0^t F_{Y_2}(t-\delta(u)) F_{Y_1}(t-\delta(v))\eta(u,v)dudv,
\end{eqnarray*}
where
\begin{eqnarray}
\eta(u,v)&=&f_{X_1}(u)f_{X_2}(v)-f_{X_1}(v)f_{X_2}(u)\nonumber
\\&=&-\eta(v,u).\label{ews30}
\end{eqnarray}
Further, $\Delta_4(t)$ can equivalently be written as
\begin{eqnarray}
 \Delta_4(t)&=&F_{H_2}(t)\int\limits_0^t\int\limits_{0}^v F_{Y_2}(t-\delta(u)) F_{Y_1}(t-\delta(v))\eta(u,v)dudv\nonumber
 \\&&+F_{H_2}(t)\int\limits_0^t\int\limits_{v}^t F_{Y_2}(t-\delta(u)) F_{Y_1}(t-\delta(v))\eta(u,v)dudv\nonumber
 \\&=&F_{H_2}(t)\int\limits_0^t\int\limits_{0}^v F_{Y_2}(t-\delta(u)) F_{Y_1}(t-\delta(v))\eta(u,v)dudv\nonumber
 \\&&+F_{H_2}(t)\int\limits_0^t\int\limits_{0}^u F_{Y_2}(t-\delta(u)) F_{Y_1}(t-\delta(v))\eta(u,v)dvdu\nonumber
 \\&=&F_{H_2}(t)\int\limits_0^t\int\limits_{0}^v F_{Y_2}(t-\delta(u)) F_{Y_1}(t-\delta(v))\eta(u,v)dudv\nonumber
 \\&&+F_{H_2}(t)\int\limits_0^t\int\limits_{0}^v F_{Y_2}(t-\delta(v)) F_{Y_1}(t-\delta(u))\eta(v,u)dudv\nonumber
 \\&=&F_{H_2}(t)\int\limits_0^t\int\limits_{0}^v \zeta_t(u,v)\eta(u,v)dudv,\label{ews31}
\end{eqnarray}
where
$$\zeta_t(u,v)=F_{Y_2}(t-\delta(u)) F_{Y_1}(t-\delta(v))-F_{Y_2}(t-\delta(v)) F_{Y_1}(t-\delta(u)),$$
and the last equality follows from (\ref{ews30}). Since $\delta(u)$ is increasing in $u\geq0$, and $Y_1\leq_{rhr}\;(\geq_{rhr})Y_2$ we have, for $0\leq t-\delta(v)\leq t-\delta(u)$,
$$\frac{F_{Y_2}(t-\delta(u))}{F_{Y_1}(t-\delta(u))}\geq\;(\leq)\frac{F_{Y_2}(t-\delta(v))}{F_{Y_1}(t-\delta(v))},$$
or equivalently,
\begin{eqnarray}
\zeta_t(u,v)\geq\;(\leq)\;0.\label{ews32}
\end{eqnarray}
Again, $X_1\leq_{lr}(\geq_{lr})X_2$ gives that, for $0\leq u\leq v\leq t$,
\begin{eqnarray}
\eta(u,v)\geq\;(\leq)\;0.\label{ews33}
\end{eqnarray}
Thus, on using (\ref{ews32}) and (\ref{ews33}) in (\ref{ews31}) we have $\Delta_4(t)\geq 0,$ and hence $Q^p_1\leq_{st}Q^p_2$.$\hfill \Box$
\\\hspace*{0.3 in}One example of the above theorem is given below.
\begin{example}
 We consider $X_1,X_2,Y_1$ and $Y_2$ same as in Example~\ref{pexs00}.
  Clearly, $X_1\leq_{lr}X_2$ and $Y_1\leq_{rhr}Y_2$.
 Assume that, for all $u\geq 0$, $\omega(u)=\ga(u)=au$, where $0< a\leq 1$.
 Then, all the conditions given in Theorem~\ref{thw67} are satisfied, and hence $Q^p_1\leq_{st}Q^p_2$.$\hfill \Box$
\end{example}
\section{Conclusions}
\hspace*{0.3 in}In this note we have studied some allocation problems in connection with general standby system. 
We have discussed three different models of one or more standby components. 
In each model we compare different series (resp. parallel) systems which are generated through different allocation strategies of standby components.
These comparisons are made with respect to the usual stochastic and the stochastic precedence orders.
Since, hot and cold standby systems are the particular cases of general standby system, thus, our discussed results generalize the existing results 
(which are connected with the hot and cold standby systems) available in the literature. 
Such a study is meaningful because it might help the design engineers to decide the best allocation strategy in order to get the optimal system depending on the underlying situation.
 \section*{Acknowledgements}
 \hspace*{0.3 in}The authors are thankful to Dr. S. P. Mukherjee, retired Centenary professor of the Dept. of Statistics, Calcutta University for some stimulating discussion.
  Financial support from Council of Scientific and Industrial Research, New Delhi (Grant No. $09/921(0060)2011$-EMR-I) is sincerely
 acknowledged by Nil Kamal Hazra.

\end{document}